\documentclass[prc,twocolumn,showpacs,preprintnumbers,amsmath,amssymb,floatfix]{revtex4}

\usepackage{graphicx}
\usepackage{dcolumn}
\usepackage{bm}
\usepackage{enumerate}

\usepackage[normalem]{ulem}  
\usepackage[dvips]{color}

\begin{document}


\title{Finite size effects in isobaric ratios}

\author{S.R. Souza$^{1,2}$}
\author{M.B. Tsang$^3$}
\affiliation{$^1$Instituto de F\'\i sica, Universidade Federal do Rio de Janeiro
Cidade Universit\'aria, \\CP 68528, 21941-972, Rio de Janeiro, Brazil}
\affiliation{$^2$Instituto de F\'\i sica, Universidade Federal do Rio Grande do Sul,\\
Av. Bento Gon\c calves 9500, CP 15051, 91501-970, Porto Alegre, Brazil}
\affiliation{$^3$National Superconducting Cyclotron Laboratory, and Joint Institute for Nuclear Astrophysics,\\ Michigan State University, East Lansing, Michigan 48824, USA}

\date{\today}

\begin{abstract}
The properties of isobaric ratios, between nuclei produced in the same reaction, are investigated using the canonical and grand-canonical statistical ensembles.
Although the grand-canonical formulae furnish a means to correlate the ratios with the liquid drop parameters, finite size effects make it difficult to obtain their  actual values from fitting nuclear collision data.
\end{abstract}

\maketitle

Central collisions between nuclei at energies starting at a few tens of MeV per nucleon lead the system to stages at which nuclear matter is hot and compressed \cite{momentumDependentPotBaoAnLi,AMD1996,XLargeSystems,exoticDens}.
If the collision is violent enough, the subsequent expansion drives it  to appreciably low densities and the nuclear disassembly takes place     
\cite{momentumDependentPotBaoAnLi,AMD1996,XLargeSystems,exoticDens}.
The properties of nuclear matter under these extreme conditions, i.e. its nuclear equation of state (EOS), play a very important role in ruling the dynamics during these stages \cite{exoticDens,reviewBaoAnLi2008,PawelIsospintTranspRatio2011,isoscalingAMD2003,TanisoDiff,constraintsDensityDepSymEnergMSU2009}.
Thus, the study of the outcome of these reactions provides important insight into the EOS.

Since the symmetry energy provide smaller contributions to the equation of state, information about symmetry energy is often obtained through ratios such as the isoscaling ratios constructed using the yields of nuclei of mass and atomic numbers $A$ and $Z$, respectively, produced in similar reactions, labeled `1' and `2',
$Y^{ (1)}(A,Z)/Y^{(2)}(A,Z)$ \cite{isoscaling3}. 
Isoscaling has been found to be closely related to the difference between the chemical potentials associated with the two systems.
Thus it allows one to exploit its relationship with the symmetry energy  \cite{isoscaling3,isoscWolfgangBotvina1}, although precise conclusions require very careful analyses \cite{isoMassFormula2008,isotemp,SubalSymEnergy,isoSMMTF,RadutaIsoSym1,RadutaIsoSym2}.

Recently, it has been suggested by the authors of Ref.\ \cite{MaScaling2011} that important information on the liquid drop parameters, including the difference in the chemical potential of neutrons and protons, may be obtained by studying isobaric ratios between nuclei produced in the same reaction:

\begin{equation}
R(I,I',A)=\frac{Y(A,I)}{Y(A,I')}\;,
\label{eq:defratio}
\end{equation}

\noindent
where $I\equiv N-Z$.
These authors offered an interpretation based on a modified Fisher model \cite{FisherModel1967} to the different correlations studied in that work.
More specifically, the yields $Y(A,I)$ were calculated from this model and several manipulations, involving different ratios, were used to extract temperature dependent liquid drop model parameters.
The results were compared to the projectile fragmentation data reported in Ref.\ \cite{MaMSUdata1}.

In this work, we resort to the standard grand-canonical and canonical statistical approaches \cite{BettyPhysRep2005} to investigate the extent to which reliable information may indeed be obtained from this analysis.
Models based on these treatments have been extensively used in the last decades and have been very successful in describing many features of nuclear collisions \cite{BettyPhysRep2005,Bondorf1995}.
In the framework of the grand-canonical approach, the yields $Y(A,I)$ reads:

\begin{eqnarray}
\label{eq:yai}
Y(A,I)&=&\frac{g_{A,Z}V_fA^{3/2}}{\lambda_T^3}e^{-[f_{A,Z}(T)-\mu_pZ-\mu_nN]/T}\\
         &=&\frac{g_{A,I}V_fA^{3/2}}{\lambda_T^3}e^{-[f_{A,I}(T)-\frac{\mu_p+\mu_n}{2}A-\frac{\Delta\mu}{2} I]/T}\;,\nonumber
\end{eqnarray}

\noindent
where $g$ stands for the spin degeneracy (taken as unit),  $\lambda_T=\sqrt{2\pi\hbar^2/m_nT}$ is the thermal wavelength, $m_n$ denotes the nucleon mass, $\mu_p$ ($\mu_n$) represents the proton (neutron) chemical potential, $\Delta\mu=\mu_n-\mu_p$, $f_{A,Z}(T)$ is the Helmholtz free energy associated with the fragment, and $T$ is the breakup temperature.
The free volume reads $V_f=\chi V_0$, where $V_0$ is the source's volume at normal density and we use $\chi=2$ throughout this work.
Upon inserting Eq.\ (\ref{eq:yai}) into Eq.\ (\ref{eq:defratio}), one finds:

\begin{equation}
\ln R(I,I',A)=\frac{\Delta\mu}{2T}(I-I')-\frac{[f_{A,I}(T)-f_{A,I'}(T)]}{T}\;.
\label{eq:rii}
\end{equation}

In order not to obscure the essential points of the present analysis, we adopt a simple prescription for the different contributions to the Helmholtz free energy:

\begin{equation}
f_{A,I}(T)=-B_{A,I}+f^*_{A,I}(T)-\frac{a_c}{(1+\chi)^{1/3}}Z^2_I/A,
\label{eq:feterms}
\end{equation}

\noindent
with $Z_I=(A-I)/2$.
More specifically, the last term in the above equation corresponds to the Wigner-Seitz correction to the Coulomb energy \cite{smm1} and  the binding energy

\begin{equation}
B_{A,I}=a_vA-a_sA^{2/3}-a_{\rm sym}I^2/A-a_cZ^2_I/A^{1/3}\;,
\label{eq:fe}
\end{equation}

\begin{figure}[tbh]
\includegraphics[width=8.5cm,angle=0]{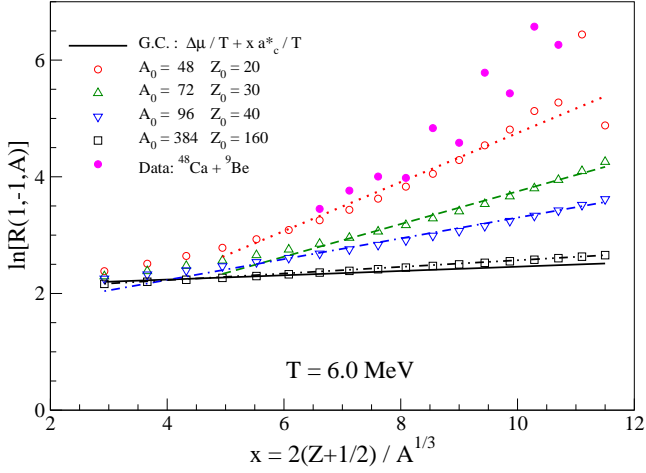}
\caption{\label{fig:MaR11} (Color online) isobaric ratios calculated for Odd A nuclei and for different system sizes in the framework of the Canonical ensemble at $T=6.0$ MeV.
The fit to the linear behavior shown in the figure gives $\Delta\mu/T=0.55$, 0.95, 1.52 and 2.00, while $a^*_c/T=0.420$, 0.280, 0.178 and 0.057 for the $(48,20)$, $(72,30)$, $(96,40)$ and $(384,160)$ nuclei, respectively.
Results obtained with the grand-canonical ensemble (solid line) as well as ratios calculated with the experimental data of Ref.\ \cite{MaMSUdata1} (full circles symbols) are also shown.}
\end{figure}

\noindent
is one of the simple formulae used in Ref.\ \cite{ISMMmass}, whose parameters read: $a_v=15.8$ MeV, $a_s=18.0$ MeV, $a_{\rm sym}=23.5$ MeV, and $a_c=0.72$ MeV.
The internal Helmholtz free energy is the same employed in Refs.\ \cite{smm1,isoMassFormula2008}

\begin{equation}
f^*_{A,I}(T)=-\frac{T^2}{\epsilon_0}A+\beta_0A^{2/3}\left[\left(\frac{T_c^2-T^2}{T_c^2+T^2}\right)^{5/4}-1\right]\;,
\label{eq:feint}
\end{equation}

\noindent
for $A\ge 5$.
It should be noted that this simple version of $f^*$ does not depend on $I$.
The parameters, are the same used in Ref.\ \cite{isoMassFormula2008}, i.e. $T_c=18.0$ MeV, $\beta=18.0$ MeV, and $\epsilon_0=16.0$ MeV.
Thus, Eq.\ (\ref{eq:rii}) becomes:

\begin{eqnarray}
\ln R(I,I',A)&=&\frac{\Delta\mu}{2T}(I-I')-\frac{a^*_c}{TA^{1/3}}(Z^2_I-Z^2_{I'})\nonumber\\
&+&\frac{a_{\rm sym}}{TA}(I^2-I'^2)\;,
\label{eq:rii2}
\end{eqnarray}

\noindent
where we have introduced $a^*_c=a_c[1-1/(1+\chi)^{1/3}]$.

The equivalent formula derived in Ref.\ \cite{MaScaling2011} differs from this one by (small) terms involving the entropy of mixing.
This small difference is slightly enhanced by factors associated with the pairing term of the binding energy, which we neglect in this work but is considered in Ref.\ \cite{MaScaling2011}.
However, since we confine our analysis to odd $A$, and focus on $I=1$ and $I'=-1$, all these terms vanish and $\ln R(1,-1,A)$ read:

\begin{equation}
\ln R(1,-1,A)=\frac{\Delta\mu}{T}+\frac{2a^*_c}{T}\frac{(Z+1/2)}{A^{1/3}}\;,
\label{eq:r11}
\end{equation}

\noindent
 where $Z=(A-1)/2$.
The last term differs from that given by Eq.\ (8) of Ref.\ \cite{MaScaling2011} because those authors write the Coulomb energy proportional to $Z(Z-1)$ whereas we use $Z^2$.

Figure \ref{fig:MaR11} shows $R(1,-1,A)$, calculated for Odd A nuclei, in both the grand-canonical and canonical approaches, at $T=6.0$ MeV.
The grand-canonical yields are obtained by determining $\mu_p$ and $\mu_n$ from the constraints $A_0=\sum_{A,Z}AY_{A,Z}$ and $Z_0=\sum_{A,Z}ZY(A,Z)$, using $Y(A,Z)$ given by Eq.\ (\ref{eq:yai}).
Then, $R(1,-1,A)$ is calculated directly from Eq.\ (\ref{eq:r11}).
The results are represented in this figure by the solid line.
By construction, it  gives  $a_c^*=0.2208$ MeV  (the slope $a^*_c/T=0.0368$)
\footnote{One should note that $a^*_c$ corresponds to the Coulomb coefficient including the Wigner-Seiz correction and, for this reason, $a_c$ is multiplied by the factor
$1-1/(1+\chi)^{1/3}=0.3068$, for $\chi=2$.}.
Owing to the exponential relationship between the chemical potentials and the system size, $\Delta\mu$ varies within 1\% in the considered mass range.
Since the variation is very small,  we use the $^{\rm 48}$Ca value $\Delta\mu=12.5458$ MeV ($\Delta\mu/T=2.091$) in all the grand-canonical plots in this work.

\begin{figure}[tbh]
\includegraphics[width=8.5cm,angle=0]{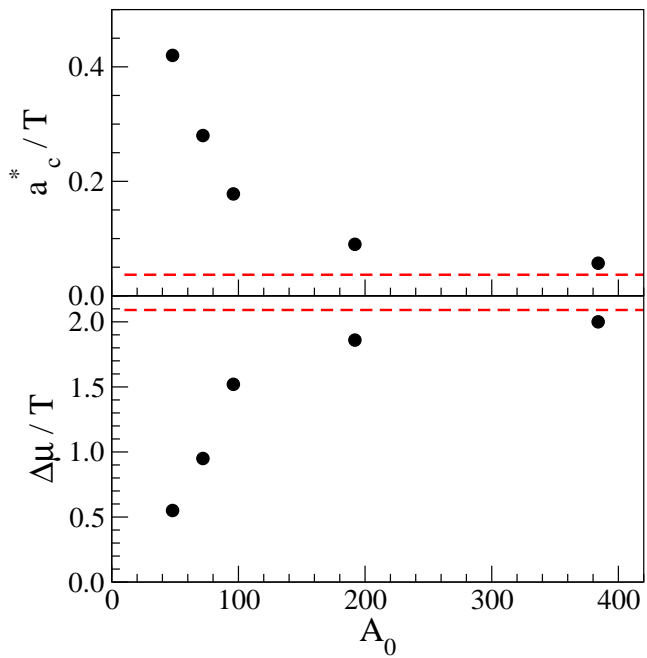}
\caption{\label{fig:parsA} (Color online) Size dependence of the fit parameters obtained with the canonical model.
The dashed lines show the values predicted by the grand-canonical approach, i.e. $a^*_c/T=0.0368$ and $\Delta\mu/T=2.091$.
For details, see the text.}
\end{figure}

The results obtained with the canonical ensemble for different sources $(A_0,Z_0)$, i.e. $(48,20)$, $(72,30)$, $(96,40)$, and $(392,160)$, are also shown in Fig.\ \ref{fig:MaR11}.
The linear dependence predicted by the grand-canonical formula, Eq.\ (\ref{eq:r11}),  gives way to curve lines which become more pronounced with smaller source size.
The linear $x$ dependence predicted by Eq.\ (\ref{eq:r11}) is a fairly accurate representation of the actual behavior only for $x\gtrsim 6$, although it should also be valid for smaller $x$ values.
Furthermore, the fit parameters are strongly dependent on the system size and lead to values of $\Delta\mu/T$ and $a^*_c/T$ appreciably different from the correct values.
They seem to converge to the asymptotic values (predicted by the grand-canonical ensemble) only in the limit of very large systems.
This is illustrated by the symbols displayed in Fig.\ \ref{fig:parsA}, which shows $a^*_c/T$ and $\Delta\mu/T$ as a function of the source size $A_0$.

This discrepancy is expected since, as discussed, for instance, in Refs.\ \cite{BettyPhysRep2005,grandCanonicalBotvina1987}, finite size effects affect the yields predicted by the canonical ensemble so that it should be equivalent to the grand-canonical approach only in the limit when $A_0\rightarrow\infty$.
Therefore, although the experimental $R(1,-1,A)$ ratios, also shown in Fig.\ \ref{fig:MaR11}, exhibit an approximate linear behavior for $x\gtrsim 6$, as asserted in Ref.\ \cite{MaScaling2011}, the values of the parameters obtained from this analysis are not clearly connected to the mass formula since the sources formed in actual experiments are not large enough to allow the finite size effects to be neglected.
The projectile fragmentation data of $^{48}$Ca~+~$^9$Be, which are depicted by the full circles in Fig.\ \ref{fig:MaR11},  correspond to one of the systems studied in Ref.\ \cite{MaScaling2011}.
Note that the actual data do not exhibit a strictly linear trend as predicted by Eq.\ (\ref{eq:r11}) and in Ref.\ \cite{MaScaling2011}.
The staggering of the data points are not reproduced by any models and are very likely due to the deexcitation of the primordial hot fragments.
Ignoring this, the data follow the trend of the models with larger curvatures and therefore larger slope giving rise to slope parameter of $a^*_c/T=0.75$ and the offset parameter of  $\Delta\mu/T=-1.7$.
From the present discussion, the extracted parameters are expected to be strongly affected by finite size effects and, therefore, their physical interpretation is seriously compromised.
It should be mentioned that we observe the same qualitative behavior in the $^{40}$Ca~+~$^9$Be and $^{58}$Ni~+~$^9$Be projectile fragmentation.
We did not study the $^{64}$Ni~+~$^9$Be system as there are very few mirror nuclei produced in this case.
It does not affect our discussion as the bottom line, i.e. the role played by finite size effects, does not depend on any particular source composition.
Nevertheless, to avoid effects from different neutron to proton composition of the source, those used in the calculations in this work have the same $N/Z$ ratios.

\

\bigskip

\

In conclusion, the linear behavior observed in the isobaric ratios studied in Ref.\ \cite{MaScaling2011}, for $2Z/A^{1/3}\gtrsim 6$, in the framework of a modified Fisher model can also be explained in terms of the standard canonical and grand-canonical statistical ensembles.
However, the latter strongly suggest that finite size effects prevent one from obtaining precise information on the nuclear properties from such isobaric ratios.
More specifically, the parameters obtained in such analysis converge very slowly to the asymptotic values as a function of the system size.
The effects are negligible only for system sizes which are much larger than those actually formed in the experiments.
Thus, our results strongly suggest that it is not possible to safely ascribe a physical meaning to the parameters obtained from such analysis.
 Furthermore, finite size effects should not be neglected in the comparison of data to observables obtained from the modified Fisher Model.

\begin{acknowledgments}
We would like to acknowledge CNPq,  FAPERJ BBP grant, CNPq-PROSUL, FAPERGS,  the joint PRONEX initiatives of CNPq/FAPERJ under
Contract No.\ 26-111.443/2010 and CNPq/FAPERGS, for partial financial support.
This work is supported by the US National Science Foundation under Grant No. PHY-1102511 and No. PHY-0822648.
\end{acknowledgments}

\bibliography{manuscript}

\end{document}